\documentstyle[prl,aps,twocolumn]{revtex}
\begin{document}

\def\r{{\bf r}}
\def\be{\begin{equation}}
\def\ee{\end{equation}}

\title{Correlation induced collapse of many-body systems with zero-range
potentials}

\author{D.V.~Fedorov \and A.S.~Jensen}
\address{Institute of Physics and Astronomy, University of Aarhus,
DK-8000 Aarhus C, Denmark}

\draft

\maketitle

\begin{abstract}
The zero-range potential is customarily employed in various mean-field
calculations of many-body systems in atomic and nuclear physics within,
correspondingly, Gross-Pitaevskii and Skyrme-Hartree-Fock approach.
We argue, however, that a many-body system with zero-range potentials is
unstable against clusterization into collapsed three-body subsystems.
We show that neither the density dependence of the potential nor an
additional repulsive three-body potential can prevent this unexpected
correlational collapse if the potentials are of zero range. Therefore
the zero-range potential can only be used in many-body calculations
where all three-body correlations are explicitly excluded.

\end{abstract}


\narrowtext

\section{Introduction}

\subsection{Many-body systems with zero-range interaction}

The zero-range potential is a very useful concept for physical problems
with a distinct separation of scales \cite{Demkov}. If, say, the typical
wavelength of a particle in a potential well is much larger than the
range of the potential the latter can be approximated by a zero-range
(pseudo)potential $V_0$ which acts on the particle's wavefunction
$\psi({\bf r})$ as \cite{Demkov,Tsurumi}

\begin{equation}\label{zrp}
V_0 \psi(\r) =
t_0\delta({\bf r})\frac{\partial}{\partial r}r\psi({\bf r}) \; ,
\end{equation}
where the strength of the potential $t_0$ is expressed in
terms of the mass $m$ and the $s$-wave scattering length $a$
of the particle\footnote{$k\cot(\delta_0) =1/a+O(k^2)$.} as
$t_0=-4\pi\hbar^2a/m$. For a well-behaving $\psi({\bf r})$ the operator
$\frac{\partial}{\partial r}r$ can be replaced with unity.

The zero-range potentials are widely used in different fields of
physics as a practically convenient form of the effective interaction.
In particular it is routinely employed in various Hartree-Fock
calculations in atomic \cite{Tsurumi,esry,stoof,Pitaevskii},
nuclear \cite{colo,ham99,ring00,dob00,naz00}, and astro-physics
\cite{nstar}.  The Hamiltonian of a many-body system is approximated
by an operator\footnote{The full Skyrme potential includes also
momentum-dependent terms which, however, are not important for the
collapse discussed here.} \cite{Tsurumi}

\begin{equation}\label{ham}
H=\sum_{i=1}^N\left(
-\frac{\hbar^2}{2m}{\bf \nabla}_i^2
\right)
+\frac{1}{2}\sum_{i\ne j} t_0\delta(\r_i-\r_j) \;,
\end{equation}
where $\r_i$ is the coordinate of the $i$-th particle and $N$ is the
total number of particles. If the system is exposed to an external field
a corresponding term is added to the  Hamiltonian.

The Hartree-Fock approximation to the wavefunction of a system of
bosons leads then to the Gross-Pitaevskii energy functional \cite{G-P},
\be\label{gp}
E_B=\int{d\r}\left(
\frac{\hbar^2}{2m}\tau_B(\r)+\frac{1}{2}t_0n_B^2(\r)
\right),
\ee
where $\tau_B(\r)$ is the kinetic energy density and $n_B(\r)$ is the
particle density of the system of bosons.  For a system of fermions one
obtains instead the Skyrme-Hartree-Fock energy functional \cite{Brink}
which, taking  symmetric nuclear matter as an example, can be written
in terms of the corresponding fermionic densities $\tau_F$ and $n_F$ as
\be\label{shf}
E_F=\int{d\r}\left(
\frac{\hbar^2}{2m}\tau_F(\r)+\frac{3}{8}t_0n_F^2(\r)
\right).
\ee

The ground state wavefunction of the Hamiltonian (\ref{ham}) minimizes
(within the Hartree-Fock approximation) the corresponding energy
functional.

\subsection{Mean-field collapse of a many-body system
with attractive zero-range interaction}

The kinetic energy $\tau_F$ of symmetric nuclear matter can be expressed
in terms of the Fermi momentum, $k_F=(3\pi^2n_F/2)^{1/3}$, as \cite{SJ}
\be
\tau_F=\frac{2}{3\pi^2}\frac{3}{5}k_F^5=
\frac{3}{5}\left(\frac{3\pi^2}{2}\right)^{2/3}n_F^{5/3} \;.
\ee
Inserting this in Eq.(\ref{shf}) shows that for large $n_F$ the potential
energy term, proportional to $n_F^2$, prevails over the kinetic energy
term, proportional to $n_F^{5/3}$. Therefore in the case of attraction,
$t_0<0$, the degeneracy pressure of the fermionic gas can not withstand
the attraction and the system collapses: the minimum of the energy,
$E_0=-\infty$, is reached at infinitely high density $n_F=\infty$.

Since there is no degeneracy pressure in a system of bosons the collapse
in this case is even more severe \cite{Salasnich}.

The mean field collapse, however, can be easily removed within the
same approach by extending the Hamiltonian (\ref{ham}) with either a
repulsive three-body zero-range potential
\be\label{tbf}
W_3=\sum_{i>j>k}t_3\delta(\r_i-\r_j)\delta(\r_i-\r_k)\;,
\ee
or a density dependent zero-range potential
\be\label{ddf}
W_3=\sum_{i>j}t_3n^\alpha(\r_i)\delta(\r_i-\r_j)\;,
\ee
which results in an additional term in the energy functional proportional
to $n^{3}$ or $n^{\alpha+2}$. The energy functionals (\ref{gp}) and
(\ref{shf}) acquire then a well-defined minimum at a finite saturation
density \cite{Brink}.

This mean-field collapse can also occur for some finite range
interactions, but an appropriate density dependence can in this
case also remove the collapse \cite{khoa93}.

\subsection{Correlational collapse of a many body system with zero-range
potentials}

Curiously enough the rigorous solution of a three-body problem with
zero-range potentials also exhibits a collapse known as the Thomas
effect \cite{thomas}. However, this is a different type of collapse,
where infinitely many bound states appear with exceedingly large
binding energies and exceedingly small spatial extension.  The many
body system with Hamiltonian (\ref{ham}) is therefore also subjected to
this non-mean-field {\em correlational} collapse -- clusterization into
collapsed three-body subsystems. Rather surprisingly this undesirable
feature of the Hamiltonian (\ref{ham}) seems to have been unnoticed
so far.

The additional potential $W_{3}$ removes the {\em mean-field} collapse of
the many-body system. But unless it also removes the {\em correlational}
collapse, the Hamiltonian (\ref{ham}) will still have this unpleasant
property of not having a finite ground state energy.  The many on-going
investigations would then be disturbingly close to a divergence, where
even a small admixture of three-body correlations might influence the
solution by a substantial and uncontrollable amount.

We shall show in this letter that neither the three-body zero-range
potential (\ref{tbf}) nor the density dependent zero-range potential
(\ref{ddf}) is actually able to remove the collapse of the many-body
system into collapsed three-body clusters.

\section{Three-body system with zero-range potentials}

\subsection{Hyper-spheric coordinates}

Let us first introduce the hyper-spheric coordinates $\{\rho,\Omega_i\}$
suitable for a description of a three-body system.  If $m_{i}$ and ${\bf
r}_{i}$ refer to the $i$-th particle then the hyper-radius $\rho $ and the
hyper-angle $\alpha _{i}$ are defined in terms of the Jacobi coordinates
${\bf x}_{i}$ and ${\bf y}_{i}$ as \cite{RR}
\begin{eqnarray}
&&{\bf x}_{i}= \sqrt{\frac{1}{m}
\frac{ m_{j}m_{k}}{m_{j}+m_{k}}}({\bf r}_{j}-{\bf r}_{k}) \;,\label{hyp} \\
&&{\bf y}_{i} = \sqrt{\frac{1}{m}
\frac{ m_{i}(m_{j}+m_{k})}{m_{i}+m_{j}+m_{k}}}\left( {\bf r}_{i}-\frac{m_{j}
{\bf r} _{j}+m_{k}{\bf r}_{k}}{m_{j}+m_{k}}\right) \;,  \nonumber \\
&&\rho \sin (\alpha _{i}) = x_i \;,\;
\rho \cos (\alpha _{i}) = y_i \;, \nonumber
\end{eqnarray}
where $\{i,j,k\}$ is a cyclic permutation of \{1,2,3\} and $m$ is an
arbitrary mass scale.  The set of angles $\Omega _{i}$ consists of the
hyper-angle $\alpha _{i}$ and the four angles ${\bf x}_{i}/|{\bf x}_{i}|$
and ${\bf y}_{i}/|{\bf y}_{i}|$. The kinetic energy operator $T$ is
defined as
\begin{eqnarray}
&&T=T_{\rho }+\frac{\hbar ^{2}}{2m\rho ^{2}}\Lambda ^{2}\;,\nonumber \\
&&T_{\rho } =-
\frac{\hbar ^{2}}{2m}\left( \rho ^{-5/2}\frac{\partial ^{2}}{ \partial \rho
^{2}}\rho ^{5/2}-\frac{1}{\rho ^{2}}\frac{15}{4}\right)\;,  \label{def} \\
&&\Lambda ^{2} =-\frac{1}{\sin (2\alpha _{i})}\frac{\partial ^{2}}{\partial
\alpha _{i}^{2}}\sin (2\alpha _{i})-4+\frac{l_{x_{i}}^{2}}{\sin ^{2}(\alpha
_{i})}+\frac{l_{y_{i}}^{2}}{\cos ^{2}(\alpha _{i})}\;,  \nonumber
\end{eqnarray}
where ${\bf l}_{x_{i}}$ and ${\bf l}_{y_{i}}$ are the angular momentum
operators related to ${\bf x}_{i}$ and ${\bf y}_{i}$.

\subsection{Hyper-spheric adiabatic expansion}

Let us now consider a three-body system with the Hamiltonian (\ref{ham}).
We start with the general hyper-spheric adiabatic expansion \cite{fed93}
of the three-body wavefunction $\Psi$

\begin{equation}\label{hsa}
\Psi(\rho,\Omega)=\frac{1}{\rho^{5/2}}\sum_n f_n(\rho)\Phi_n(\rho,\Omega)\;,
\end{equation}
in terms of the complete basis $\Phi_n$ of the solutions of the
hyper-angular eigenvalue equation

\be \label{eigens}
\left(\Lambda^2-\lambda_n(\rho)+\frac{2m\rho ^{2}}{\hbar ^{2}}
\sum_{i=1}^{3}V_{i}(\rho ,\Omega )\right) \Phi_n(\rho ,\Omega )=0\;,
\ee
where $V_{i}$ is the potential between particles $j$ and $k$ ($i$,$j$,$k$
is the cyclic permutation of 1,2,3).  If we truncate the infinite sum in
(\ref{hsa}) we shall, according to the variational principle, obtain an
upper bound on the discrete spectrum of the system. If then the truncated
expansion provides a collapse the full wavefunction will collapse as
well. It is then sufficient to consider only the lowest term in the
expansion -- the so called {\em hyper-spheric adiabatic approximation},

\begin{equation}
\Psi(\rho,\Omega)=\frac{1}{\rho^{5/2}}f(\rho)\Phi(\rho,\Omega)\;,
\end{equation}
where the angular coordinates $\Omega$ correspond to the ''fast''
subsystem while the hyper-radius $\rho$ represents the ''slow'' subsystem.
Within the adiabatic approximation the lowest eigenvalue $\lambda(\rho)$
of the ''fast'' subsystem simply serves as the effective potential for
the ''slow'' hyper-radial subsystem,

\begin{equation}\label{rad} \left(
-\frac{\partial ^{2}}{\partial \rho ^{2}}+\frac{\lambda (\rho )+15/4}{
\rho ^{2}}-\frac{2mE}{\hbar ^{2}} \right) f(\rho )=0\;,
\end{equation}
where we have neglected one term, originating from the derivatives of the
angular functions \cite{fed93}, which is unimportant for the following
discussion of the collapse.

For zero-range potentials the Faddeev equations provide a more convenient
basis for analytic insights into the properties of the three-body system
\cite{fed93}.  The Faddeev decomposition of the angular wavefunction
$\Phi$ is
\begin{equation}
\Phi=\sum_{i=1}^{3}\frac{\phi_i}{\sin(2\alpha_i)}\;,
\end{equation}
where
the three Faddeev components $\phi_i$ satisfy the
three coupled
Faddeev equations
\begin{equation} \label{fad}
\left( \Lambda ^{2}-\lambda (\rho )\right) \frac{\phi _{i}}{\sin (2\alpha
_{i})}+\frac{2m\rho ^{2}}{\hbar ^{2}}V_{i}\Phi =0\;,\;i=1,2,3\;.
\end{equation}

The system of Faddeev equations (\ref{fad}) is equivalent to the original
Schr\"{o}dinger equation (\ref{eigens}).

\subsection{Adiabatic solutions for the zero-range potentials}

The zero-range potentials vanish identically except at the origin and we are
therefore left with the free Faddeev equations
\be
\left( -\frac{\partial ^{2}}{\partial \alpha_i^{2}}-\nu ^{2}(\rho )\right)
\phi_i(\rho,\alpha_i)=0 \; ,
\ee
where $\nu^2=\lambda+4$, and where we have restricted each of the
components of the wavefunction to $s$-waves only.  The solutions,
which obey the boundary condition $\phi_i(\rho,\frac{\pi}{2})=0$, are
\be\label{sol}
\phi_i(\rho,\alpha_i)=A_i \sin
\left[ \nu \left( \alpha_i-\frac{\pi}{2} \right) \right] \;.
\ee
For three identical particles the index $i$ can be dropped completely. For
the fermionic case we imply instead that we have a system of, say, two
neutrons and a proton in a state with spatially symmetric wavefunction.

The zero-range potential appears as a boundary condition at $\alpha =0$
\cite{Demkov}
\begin{equation}\label{bc}
\left(
\frac{1}{\alpha\Phi}\frac{\partial\alpha\Phi}{\partial\alpha}
\right)
_{\alpha =0}=\frac{\rho }{\sqrt{\mu }}\frac{1}{a} \;,
\end{equation}
where $\mu =(1/m)(m_{1}m_{2}/(m_{1}+m_{2}))$ is the reduced mass of the two
particles in units of the mass $m$, and $a$ is the scattering
length. The total angular wavefunction $\Phi$ can be expressed for
small angles within a given Jacobi system as~\cite{fed93}
\begin{equation}\label{expan}
\sin(2\alpha)\Phi(\alpha)=
\phi(\alpha)+
\frac{8}{\sqrt{3}}\alpha\phi(\frac{\pi}{3})+O(\alpha ^{2})\;.
\end{equation}
Substituting (\ref{expan}) and (\ref{sol}) into (\ref{bc}) leads to the
eigenvalue equation~\cite{fed93} for $\nu$
\begin{equation} \label{nu}
\frac{ -\nu\cos(\nu \frac{\pi }{2}) +\frac{8}{\sqrt{3}}\sin(\nu\frac{\pi}{6}
) }{ \sin(\nu \frac{\pi }{2}) }=\frac{\rho }{\sqrt{\mu }}\frac{1}{a}\;.
\end{equation}
The solution $\nu (\rho )$ of this equation defines the needed adiabatic
potential $(\nu ^{2}-1/4)/\rho ^{2}$ for the hyper-radial equation
(\ref{rad}) from which one obtains the total energy and the radial wave
function of the system.

\section{Correlational collapse of many-body system}

In the small distance region, $\rho \ll a$, the eigenvalue equation
(\ref{nu}) has an imaginary root $\nu _{0}=ib$, where $b=1.006$,
leading to an effective potential in the hyper-radial equation which
in this region behaves as $(\nu_0^2-1/4)/\rho^{2}\equiv -C/\rho ^{2}$,
where $C=1.262$. The hyper-radial equation (\ref{rad}) then becomes
\begin{equation}
\left( -\frac{\partial ^{2}}{\partial \rho ^{2}}-\frac{C}{\rho ^{2}}-\frac{
2mE}{\hbar ^{2}}\right) f(\rho )=0\;,\;\rho \ll a.
\end{equation}
The (negative) energy $E=-\hbar ^{2}\kappa ^{2}/(2m)$ is negligible
compared to the potential when the distance is sufficiently small,
$\rho \ll \kappa ^{-1}$.  The hyper-radial equation then turns into

\begin{equation}
\left( -\frac{\partial ^{2}}{\partial \rho ^{2}}-\frac{C}{\rho ^{2}}\right)
f(\rho )=0\;,
\end{equation}
which has solutions of the form $f(\rho)\sim\rho ^{n}$, where
$n=\frac{1}{2}\pm\frac{1}{2}\sqrt{1-4C}=\frac{1}{2}\pm\nu_0$. For
$C>\frac{1}{4}$ the exponent $n$ acquires an imaginary part $\pm ib$, i.e.

\begin{equation}
f(\rho )\sim \sqrt{\rho }e^{\pm ib\ln(\rho)}\;.  \label{col}
\end{equation}
This wavefunction has independent of energy infinitely many nodes at
small distances or, correspondingly, infinitely many lower lying states
at smaller distances (Thomas effect). In other words, there is actually
no finite ground state -- the three-body system collapses.

For the many-body system the energy minimum, minus infinity, is therefore
reached at the configuration, where the many-body system is clusterized
into separated three-body subsystems.  The density of the many-body system
in this configuration is then represented by a sum of delta-functions
corresponding to collapsed three-body clusters. The Hartree-Fock ansatz
for the wavefunction, however, explicitly excludes such clusterized
configurations. It is therefore safe to use the Hamiltonian (\ref{ham})
with additional potentials (\ref{tbf}) or (\ref{ddf}) as long as the
model space is restricted to the Hartree-Fock product wavefunctions.

Let us now consider the {\em extended Hamiltonian} which includes
the additional potential $W_3$ either as the three-body zero-range
potential (\ref{tbf}) or as the density dependent zero-range potential
(\ref{ddf}). Applied to a separated three-body cluster both potentials
are non-vanishing only when all three constituents are located at
the same point in space. This configuration corresponds to $\rho =0$
or $\ln(\rho)=-\infty $. These additional potentials, therefore, will
only change the boundary condition at $\rho=0$. However, independent
of the boundary condition at $\ln(\rho)=-\infty $, the infinitely many
nodes of the wave function (\ref{col}) remain and, therefore, the system
still collapses.

The repulsive zero-range potentials (\ref{tbf}) or (\ref{ddf}) need
three particles located at the same point to provide the stabilizing
contribution to the energy of the system.
Unfortunately this configuration, as we have shown, is unstable under
collapse.
If, however, this three-body collapse is somehow removed the {\em
four-body correlations} (or higher) will not destabilize the many-body
system.
Indeed the repulsive potentials (\ref{tbf}) or (\ref{ddf}) provide a
contribution to the energy, which has a higher density dependence than
that from the attractive two-body potential and hence high density
configurations will not be energetically favored.
The removal of the three-body collapse, which allows the repulsive
potentials (\ref{tbf}) or (\ref{ddf}) to contribute, will therefore also
stabilize the four and higher order correlations.

\section{Discussion}

We have shown that zero-range potentials, applied to a many-body
system in coordinate space, lead to a specific collapse of the system
driven by three-body correlations. Neither the density dependence of
the potential nor the three-body zero-range potential can remove this
correlational collapse. The Hamiltonian (\ref{ham}), perhaps extended by
the additional potential (\ref{tbf}) or (\ref{ddf}), can still produce
meaningful results, provided the allowed variational model space is
chosen consistently.  Any three-body correlation must be excluded a
priori from the wavefunction of the many-body system approximated by the
Hamiltonian (\ref{ham}). Going beyond the Hartree-Fock (anti)symmetrized
product wavefunction would immediately introduce dangerous effects of
three-body correlations.

To remove the three-body collapse one has to regularize the zero-range
potential by introducing a finite length scale $R$, which will alter
the $-C/\rho ^{2}$ behaviour of the effective potential at $\rho\sim R$.
Such regularization is automatically achieved by finite-range two-body
potentials.  The three-body system will then have approximately $\ln(a/R)$
bound states with the ground state having a finite binding energy of
the order of $\hbar ^{2}/(2mR^{2})$.

When the three-body system is regularized the correlations of higher
number of particles can be safely included in the many-body wavefunction
as the ordinary additional potentials (\ref{tbf}) or (\ref{ddf}) will
make the system stable.

\end{document}